# Multimodal Road Network Generation Based on Large Language Model

Jiajing Chen[1], Weihang Xu[2], Haiming Cao[1], Zihuan Xu[1], Yu Zhang[1], Zhao Zhang[1,3,*], Siyao Zhang[1]

1. College of Transportation Science and Engineering, Beihang University
2. College of Electronic Information Engineering, Beihang University
3. Key Laboratory of Intelligent Transportation Technology and Systems, Ministry of Education, Beijing, China

**Abstract**

With the increasing popularity of ChatGPT, large language models (LLMs) have demonstrated their capabilities in communication and reasoning, promising for transportation sector intelligentization. However, they still face challenges in domain-specific knowledge. This paper aims to leverage LLMs' reasoning and recognition abilities to replace traditional user interfaces and create an "intelligent operating system" for transportation simulation software, exploring their potential with transportation modeling and simulation. We introduce Network Generation AI (NGAI), integrating LLMs with road network modeling plugins, validated through experiments for accuracy and robustness. NGAI's effective use has reduced modeling costs, revolutionized transportation simulations, optimized user steps, and proposed a novel approach for LLM integration in the transportation field.

**1 Introduction**

Since 2022, large language models (LLMs), exemplified by ChatGPT[1], have demonstrated their formidable communicative and logical reasoning capabilities, highlighting substantial potential applications and economic impacts. This offers a promising direction for the Intelligent Transportation System (ITS). However, it must be acknowledged that LLMs still harbor fatal flaws when addressing complex issues. Taking the transportation domain as an example, they inherently lack the precise understanding of transportation issues, leading to inaccuracies and unreliability when confronting intricate problems, thus posing challenges to their integration with transportation systems.

Simultaneously, it is well recognized that traffic simulation is a common and crucial tool in transportation system research. Commonly used traffic simulation software includes TransCAD, Visum, CUBE, Vissim, SUMO, and DRACULA. The utilization of these software tools for road network modeling invariably incurs considerable temporal and cognitive investments. Therefore, considering the developmental trajectory of artificial intelligence, we envision a form of integration combining LLMs with road network modeling to reliably alleviate the troubles stemming from road network modeling.

We propose Network Generation AI (NGAI) — an integration of LLMs and road network modeling. NGAI incorporates four road network modelling methodologies, endowing LLMs with the capability to handle multimodal data input, processing, and generation during the invocation of these plugins. Through the establishment of appropriate plugin descriptors and the provision of pertinent prompts, LLMs autonomously determine the invocation of plugin names and sequence, thereby facilitating the automated completion of the road network modeling process. This fusion enables precise extraction of user requirements, reducing the learning costs and tedious work associated with road network modeling. In the subsequent chapters, we will delve into the architecture, methodologies, and case study of NGAI, demonstrating how it simplifies the road network modeling process and achieves multimodal road network modeling.

* Corresponding author: Zhao Zhang, zhaozhang@buaa.edu.cn.

## 2 Literature Review

### 2.1 Large Language Model Application

Large Language Models (LLMs) have emerged with more vertically tailored applications that meet industry-specific needs across various domains, such as medicine, sports, and finance. In the medical field, LLMs can analyze extensive medical literature and case data to provide auxiliary decision-making and guidance for physicians, thereby improving diagnostic accuracy and treatment outcomes[2][3]. In sports, LLMs can analyze athletes' physiological data, technical parameters, and competition record to offer personalized training advice to coaches and athletes, helping them enhance their competitive level[4][5]. In finance, the application of LLMs includes the automatic generation of financial reports, market trend prediction, investor sentiment analysis, and personalized financial advice. By analyzing vast financial data and market information, LLMs can produce accurate financial reports and provide advice on investment decisions and risk management. Additionally, LLMs can analyze social media and news to understand investor sentiment and market trends, assisting investors in making informed decisions[6][7]. As AI technology continues to develop, there is no doubt that LLMs will play an increasingly significant role across various domains, actively contributing to the development and progress of related industries.

Notably, scholars in the transportation field have also made contributions in integrating LLMs with transportation disciplines, actively exploring the application of LLMs in traffic planning, traffic management and control, and big data processing in transportation. For Traffic Signal Control (TSC), Lai S et al.[8] employed LLMs as decision agents for TSC to perform human-like reasoning and decision-making processes, thereby achieving effective traffic control. Experimental results demonstrate significant effectiveness, generalization capabilities, and interpretability. Liu C et al.[9] utilized the exceptional abilities of LLMs in time series analysis for traffic prediction, with comprehensive experiments proving the robust performance of ST-LLM. Ou Zheng et al.'s research[10] on LLMs in traffic safety reveals infinite possibilities for LLMs in the transportation domain. LLMs can automate accident reporting, including extracting accident information, attribution and analysis, enhancing traffic data, and conducting multisensory safety analysis.

### 2.2. Road Network Modeling

Road network modeling is a core function of transportation software and serves as the foundation for handling complex traffic tasks and analyzing traffic data[11]. Multimodal road network modeling aims to achieve a more comprehensive understanding and analysis of urban or regional transportation systems, facilitating Sustainable Urban Mobility Plans (SUMP) to optimize traffic flow, reduce congestion, improve environmental quality, and enhance travel efficiency, thus attaining more efficient and sustainable urban transportation management and planning goals[12][13].

Currently, mainstream road network modeling approaches include: Agent-Based Models (ABM) based on microsimulation[14], Four-Step models based on macroscopic modeling[15], Continuous Flow Models based on continuous modeling[16], and Machine Learning-based Prediction Models[17]. These modeling methods are suitable for different scenarios, each with its advantages and disadvantages. Overall, they require considerable temporal and cognitive investments, extensive and high-quality data support, and demand significant computational resources.

In past research, ArcGIS has made significant progress in the field of road network modeling[18], covering various aspects such as road network data models, road network analysis algorithms, spatial data visualization, and applications. The primary goal is to address various road network-related issues by analyzing and modeling the road network structure and spatial relationships[19].

In terms of road network data models, previous research has mainly focused on the construction and management of road network datasets. ArcGIS provides a rich set of tools and functionalities to support the creation and management of road network data models, such as the ArcGIS Network Analyst extension plugin. MAHINI A.S. et al.[20] introduced how to use ArcGIS for dynamic modeling, simulation, and prediction of urban growth patterns, providing decision-makers with timely information to adopt appropriate urban development policies.

Road network analysis algorithms are another key area, where efficient and accurate algorithms can be used to solve various road network optimization and routing problems. These algorithms include shortest path algorithms, network flow algorithms, minimum spanning tree algorithms, etc. ArcGIS offers a range of powerful road network analysis tools, such as shortest path analysis, service area analysis, route optimization, etc., helping users perform various road network analysis tasks. S Martín-Fernández et al.[21] explored the application of simulated annealing algorithms in GIS for solving fire station location problems, demonstrating their effectiveness in finding approximate optimal solutions in network optimization tasks.

Spatial data visualization presents complex road network data to users, aiding in understanding spatial relationships and road network structures.[22] ArcGIS provides a rich set of map-making and visualization tools, allowing users to create various types of maps, including road network maps, route maps, flow maps, etc., thereby more intuitively presenting road network data.

Furthermore, ArcGIS has a wide range of applications in road network modeling. For example, in the field of transportation planning, Moggan Motamed[23] utilized ArcGIS for traffic network modeling and simulation to evaluate the effects of traffic policies. In the field of facility location, S. Bolouri et al.[24] used ArcGIS for site selection analysis to determine the best facility location options. These application cases demonstrate the significant role of ArcGIS combined with road network modeling in addressing real-world problems.

Previous research demonstrates the application of ArcGIS in road network modeling, laying a robust foundation for future investigations. With technology continually advancing, ArcGIS is poised to achieve further breakthroughs in the field of road network modeling.

**2.3. Integration of LLM and Transportation**

With the rapid development of big data, scholars have begun to integrate LLMs with road network modeling and have achieved some results. Sumedh Rasal et al.[25] designed a model specifically for processing aerial images of road layouts. They utilized multimodal LLMs to input images and generate detailed, navigable road networks within the input images. The core innovation lies in a unique training method used to train LLMs to generate road networks as their output. This research represents a significant advancement in enhancing autonomous navigation systems, especially in road network scenarios where accurate navigation guidance is crucial.

In recent years, scholars have started to explore the combination of LLMs with Geographic Information System (GIS) platforms like ArcGIS to improve precision, efficiency, and intelligent level of road network modeling, promoting the application of GIS in traffic management and planning. These studies cover various aspects, including data enhancement, spatiotemporal data prediction, road vectorization, and traffic forecasting, with significant application potential.

LLMs possess excellent capabilities in text understanding and generation, enabling the extraction of knowledge related to road networks and geospatial information from large-scale text data. This provides new ideas and methods for constructing more accurate and comprehensive road network models. For instance, scholars can utilize LLMs for semantic understanding and inference of network topologies, better capturing the relationships between nodes and edges in road networks, and improving the accuracy of road network models.

The combination of LLMs and ArcGIS can achieve more intelligent route planning and traffic condition forecasting. LLMs can analyze vast historical path data and real-time traffic information, learning road usage patterns and traffic flow dynamics to predict future traffic conditions. Combined with the spatial analysis and visualization functions of ArcGIS, these forecast results can be intuitively presented on maps, providing important decision support for fields such as traffic management and smart city construction.

LLMs can also achieve seamless interaction between language and geospatial information. Users can query the system for road network information through natural language instructions, such as shortest paths and optimal routes, and the system can accurately understand the user's intent and provide corresponding geospatial analysis results. This linguistic interface can offer users more convenient and intuitive answers.

However, most related research to date involves single data input and fixed-form output. Recognizing the necessity and importance of utilizing LLMs to accomplish traffic modeling under multimodal inputs to reduce the temporal and cognitive costs associated with road network modeling for transportation professionals, we propose Network Generation AI (NGAI), which supports multimodal inputs and delivers personalized modeling results that meet users' expectations.

## 3 Methodology

The multimodal road network generation utilizes the integration of advanced language models and specialized plugins within the LangChain framework, capitalizing on LLMs to interpret commands in various formats and produce road network files in the General Modeling Network Specification (GMNS) format. This format supports both static and dynamic transportation planning. NGAI benefits significantly from specialized plugins capable of processing multimodal inputs like text and images into two categories: text-based and image-based road network generation tools. These plugins operate autonomously, utilizing both local and online data to fulfill precise user specifications. The harmonious integration of these components forms the innovative NGAI system, streamlining the road network modeling process.

### 3.1 Preliminaries

#### 3.1.1 LangChain

LangChain is a resilient framework designed to facilitate the development of end-to-end applications leveraging language models. It simplifies the construction process by offering a comprehensive suite of tools, components, and interfaces. LangChain adeptly manages user interactions with language models, establishing links between various components and integrating additional resources such as APIs and databases. Serving as an intermediary layer, it effectively connects user-facing applications with LLMs.

#### 3.1.2 GMNS Format

GMNS, introduced by the Zephyr Foundation, is the abbreviation for General Modeling Network Specification, which outlines a universally understandable format for sharing routable road network files, catering to both human and machine readers. GMNS is crafted for application in both multimodal static and dynamic transportation planning and operational models, aiming to enhance the collaboration among modelers by enabling the seamless exchange of tools and data sources.

The GMNS format generally consists of two CSV files: Node.csv and Link.csv. Node.csv is used to describe nodes or intersections in the traffic network and typically includes the following attributes:

- Node ID: Each node should have a unique identifier to distinguish different nodes in the network.
- Spatial location: This includes the geographical coordinates of the node, usually represented by longitude and latitude, allowing for precise positioning of the node on a map.
- Other relevant attributes: Depending on specific applications and requirements, additional attributes can be added, including node type (intersection, station, etc.), node capacity, traffic signal information, pedestrian pathways, etc.

Link.csv is used to describe road segments in the traffic network and generally includes the following attributes:

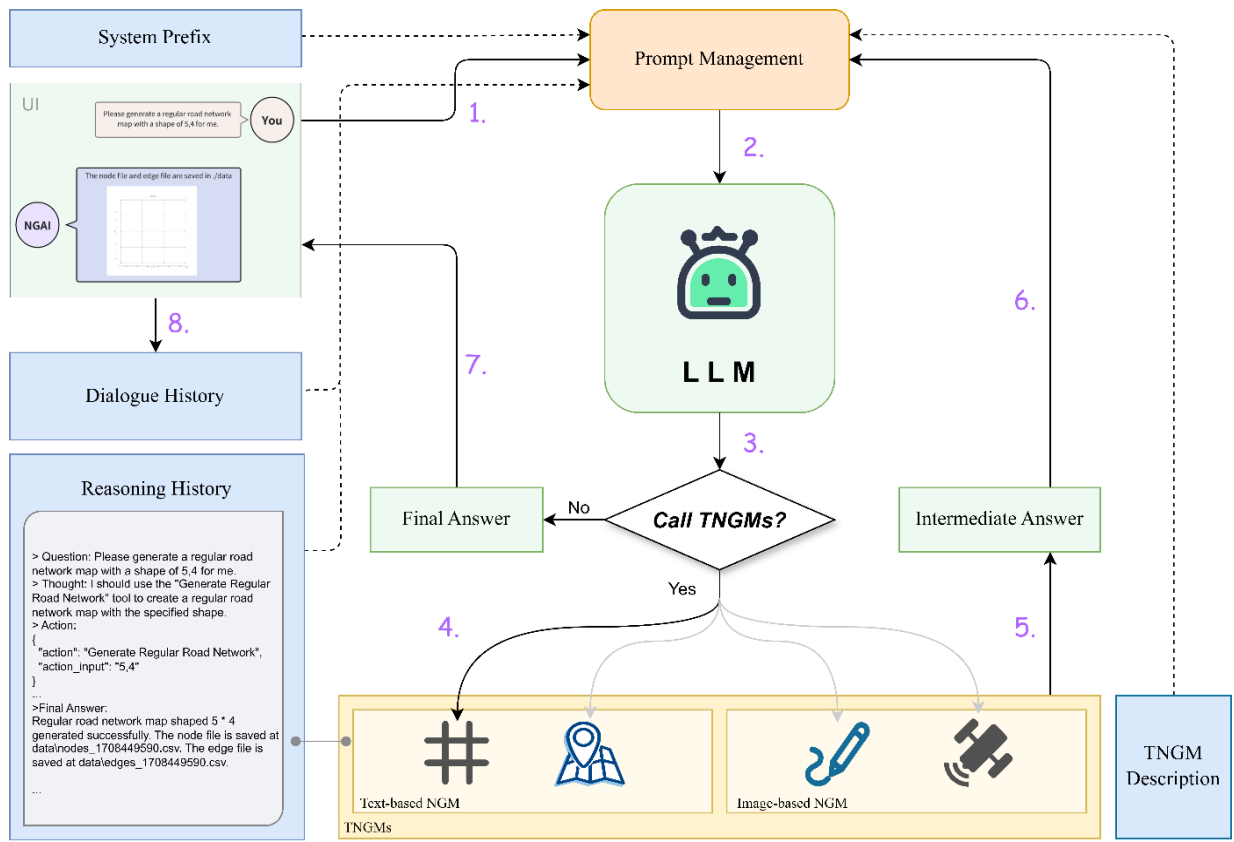

Figure 1. The Structure and Workflow of NGAI

- Link ID: Each road segment should have a unique identifier to differentiate between different roads in the network.
- Start and end nodes: Each road segment should have a unique identifier to differentiate between different roads in the network.
- Number of Lanes: The number of lanes in the road segment, used for simulating traffic flow and congestion. Default is 2.
- Geometry: This often represents the geometric information or properties of the network link. Geometry information describes the geometric characteristics of the physical path connecting two traffic network nodes, such as the shape, length, curvature of the road, etc. This is crucial for operations like traffic simulation and route planning.
- Other relevant attributes: Depending on specific applications and requirements, additional attributes can be added, including road grade, traffic signals, road names, road signs, etc.

Given the advantages, in our study, LLMs proficiently handle multimodal command inputs and produce GMNS-formatted road network files.

This format facilitates seamless integration into various traffic simulation software, enhancing accessibility and compatibility across platforms.

### 3.2 NGAI Overview

The overview of NGAI is illustrated in Figure 1. NGAI primarily consists of multimodal plugins and a large language model, interconnected through the LangChain framework. The LangChain framework empowers NGAI with functionalities such as prompt management, plugin description, and dialogue history. The plugins, developed autonomously, are categorized into Text-based Road Network Generation Tools and Image-based Road Network Generation Tools. These plugins can access local or online data sources. Upon receiving a request from the user, the LLM parses the user prompt along with the descriptors of each plugin as input, thereby determining the plugin to be invoked. Upon completion of plugin execution, the output message is returned to the LLM, which then assesses whether the user request has been fulfilled. If completed, the current cycle concludes; if not, the LLM selects the next plugin to be invoked, repeating the aforementioned process.

**3.3 Text-based Road Network Generation**

Text-based road network generation refers to the conversion of various textual information pertaining to the target road network into structured data suitable for traffic simulation, specifically in the form of GMNS files in our work. We have developed two tools: The Regular Road Network Generation Tool and the Location-specified Road Network Generation Tool. The former is dedicated to generating grid-shaped road network files. The latter, leveraging OpenStreetMap's open-source map data, can generate road network files based on user-specified locations, offering a more dynamic and customizable approach to road network generation.

**3.3.1 Regular Road Network Generation**

The Regular Road Network Generation Tool accepts textual inputs from users, leveraging LLM to extract the desired number of rows (N) and columns (M) for the target grid-shaped road network. Subsequently, the tool invokes relevant modules to process this information. These modules are designed to generate GMNS-formatted road network files while simultaneously creating visual representations of the network. The pathways of the GMNS files are ultimately returned to the LLM as output, providing users with an accessible means to view and utilize the generated road network. The principle of the Regular Road Network Generation Tool is illustrated in Figure 2.

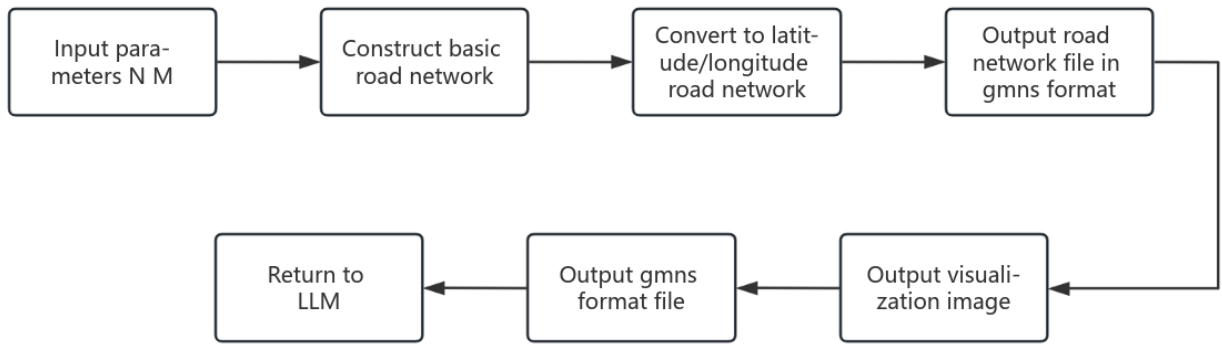

Figure 2. The internal workflow of the Regular Road Network Generation tool module

The detailed description of the workflow for the Regular Road Network Generation Tool is as follows:
- Data input: Initiated by a direct dialogue with LLM, input parameters {N,M} are acquired. Subsequently, LLM invokes functional modules to process these parameters.
- Road network construction: Initially, a rectangular road network of dimensions N ×M is constructed based on the values of N and M. The distance between adjacent nodes is set to 1. The tool then establishes the latitude and longitude of the central point for coordinate transformation, with default coordinates set at (39.125,161.567). To ensure network precision, a scale factor is introduced to specify the distance between adjacent nodes. This scale factor is an adjustable parameter with a default value of 0.004. Validation indicates that at this scale, the network error is within 500 meters, meeting user requirements.
- File generation: The LLM calls corresponding functions to draw a visual representation of the road network, offering users immediate feedback, as depicted in Figure 3. The generated GMNS-formatted road network files and visual representation are concurrently saved in a predefined directory.
- LLM output: The LLM informs the user that the road network building process is complete and provides the path to the relevant files for user retrieval.

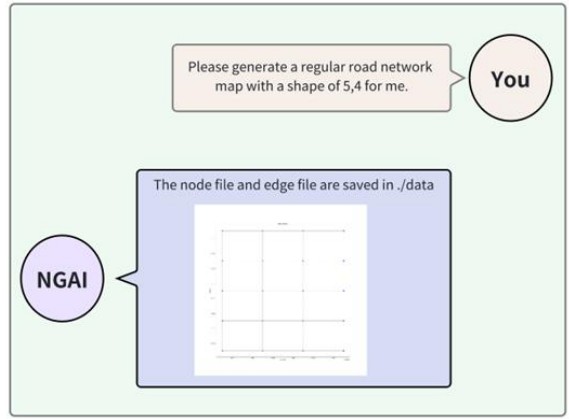

Figure 3. The interaction process between the tool and the user

**3.3.2 Location-specified Road Network Generation**

The Location-specified Road Network Generation Tool obtains the desired real-world location and scope for road network generation by parsing textual input from the user through LLM. The geographical coordinates of the target location are acquired through geocoding. The tool then retrieves osm-formatted road network files from OpenStreetMap, an online map collaboration initiative featuring open-source availability, online editing capabilities and free access to global geographic data[26], ultimately converting them into the GMNS format. Simultaneously, the tool generates a visual representation of the road network. The operational principle of the Location-specified Road Network Generation Tool is elucidated in Figure 4.

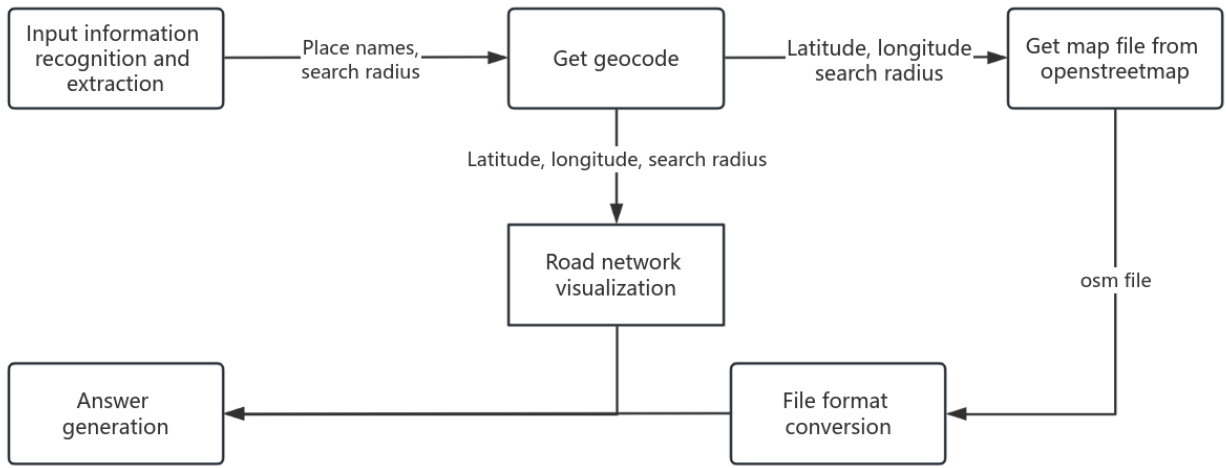

Figure 4. The internal workflow of the Location-specified Road Network Generation tool module

The workflow of the Location-specified Road Network Generation Tool is presented as follows:
- Target recognition: Leveraging the LangChain framework, LLM identifies the target location name and the scope for road network construction from the input textual information.
- Geocoding: Geocoding is a spatial encoding method based on location-based technology, providing a means to convert geolocation information described in addresses into geographical coordinates usable in Geographic Information System (GIS). The process typically involves address standardization and address matching.[27] In this tool, we achieve this process by invoking the API of the Amap, directly obtaining specific latitude and longitude coordinates based on the input location name.

- Retrieval of road network files from OpenStreetMap: Obtaining road network files from OpenStreetMap involves setting up a web crawler program, inputting location coordinates and search radius to directly download osm-formatted road network files relevant to the specified area from OpenStreetMap.
- Format conversion: The obtained road network information is in the osm file format, which is not conducive to road network modeling. In this process, the osm2gmns library in Python is utilized to convert the osm file into the required GMNS-formatted files.
- Road Network Visualization: The tool generates a visual representation of the road network based on latitude, longitude, and search radius utilizing the osmx library. In contrast to drawing from the GMNS file, this method exhibits superior temporal efficiency.
- LLM output: The LLM outputs the path of the generated GMNS file in a predefined format, facilitating user retrieval and viewing.

**3.4 Image-based Road Network Generation**

Under specific circumstances, researchers encounter a need to extract road network information from an image and transform it into structured data in the process of constructing road networks for traffic simulation. This image could be a satellite remote sensing image of a location or a hand-drawn conceptual map of a road network created by researchers. To address this requirement, we have developed two tools: The Road Network Generation Tool from Satellite Image and the Road Network Generation Tool from Hand-drawn Image. These tools employ distinct preprocessing steps to obtain a binary representation of the road network portion from the input image. Subsequently, both tools generate GMNS-formatted road network files utilizing a shared procedure.

**3.4.1 Road Network Generation from Satellite Image**

To transform a satellite image into a binary representation, we employed a U-Net model designed to extract the road network segment from the satellite image. The training dataset originates from the open-source Massachusetts Roads Dataset[28]. The training process is illustrated in Figure 5. The U-Net model, depicted in Figure 6, is a classical fully convolutional neural network architecture, particularly well-suited for image segmentation tasks[29]. Proposed by Ronneberger et al. in 2015, this model has achieved notable success, particularly in the field of medical image segmentation.

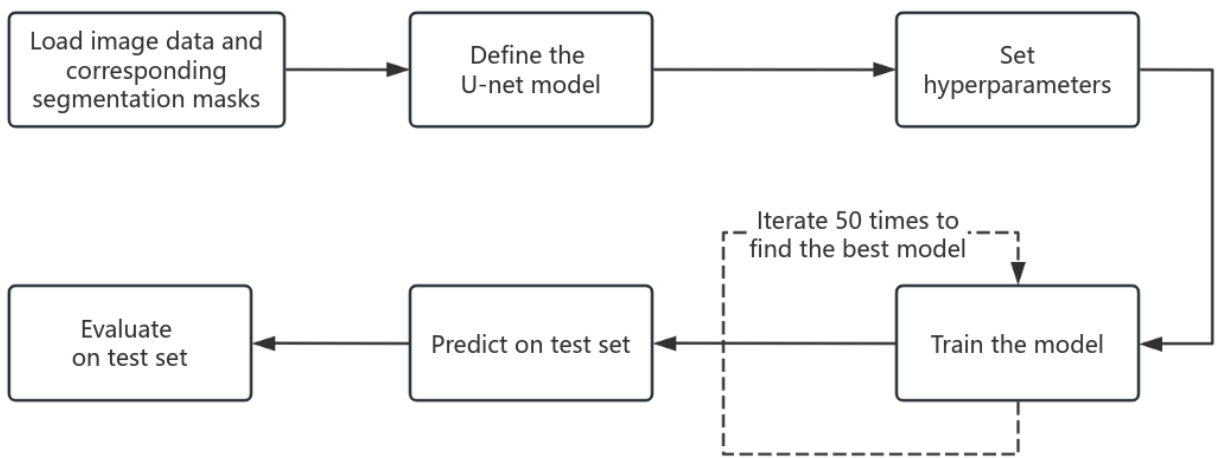

Figure 5. The process of training U-Net for extracting road networks from satellite

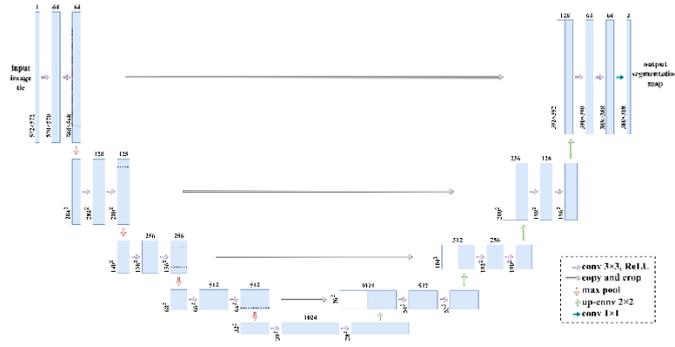

Figure 6. The architectural feature of U-Net is the integration of the encoder and decoder into a U-shaped network structure. This configuration allows the network to simultaneously leverage local details and global contextual information, thereby enhancing segmentation accuracy

In our work, the training epochs are set to 50 cycles to ensure that the model gains enough learning and adaptability during the training process. The evaluation metric employed in the model is Intersection over Union (IoU), measuring the overlap between predicted results and true labels in target detection and image segmentation. IoU provides an objective assessment of the model's performance. The IoU score is calculated using the following formula:

$$IoU = \frac{\text{Intersection of Predicted and True Regions}}{\text{Union of Predicted and True Regions}}$$

For the loss function in image segmentation tasks, the model utilizes Dice Loss, quantifying the similarity or overlap between predicted results and true labels. The computation of Dice Loss is defined by calculating the complement of the Dice coefficient between predicted results and true labels. The specific formula is as follows:

$$DiceLoss = 1 - 2 \times \frac{\text{Intersection of Predicted and True Regions}}{\text{Sum of Predicted and True Regions}}$$

Furthermore, we select the Adam optimizer and sets the learning rate to 0.00008. Through these definitions and configurations, the model is then trained and optimized to achieve accurate recognition of road networks in satellite images.

To extract the road network from the binary image, we propose an image point-line information extraction method based on Shi-Tomasi corner detection. The overall model structure is illustrated in Figure 7.

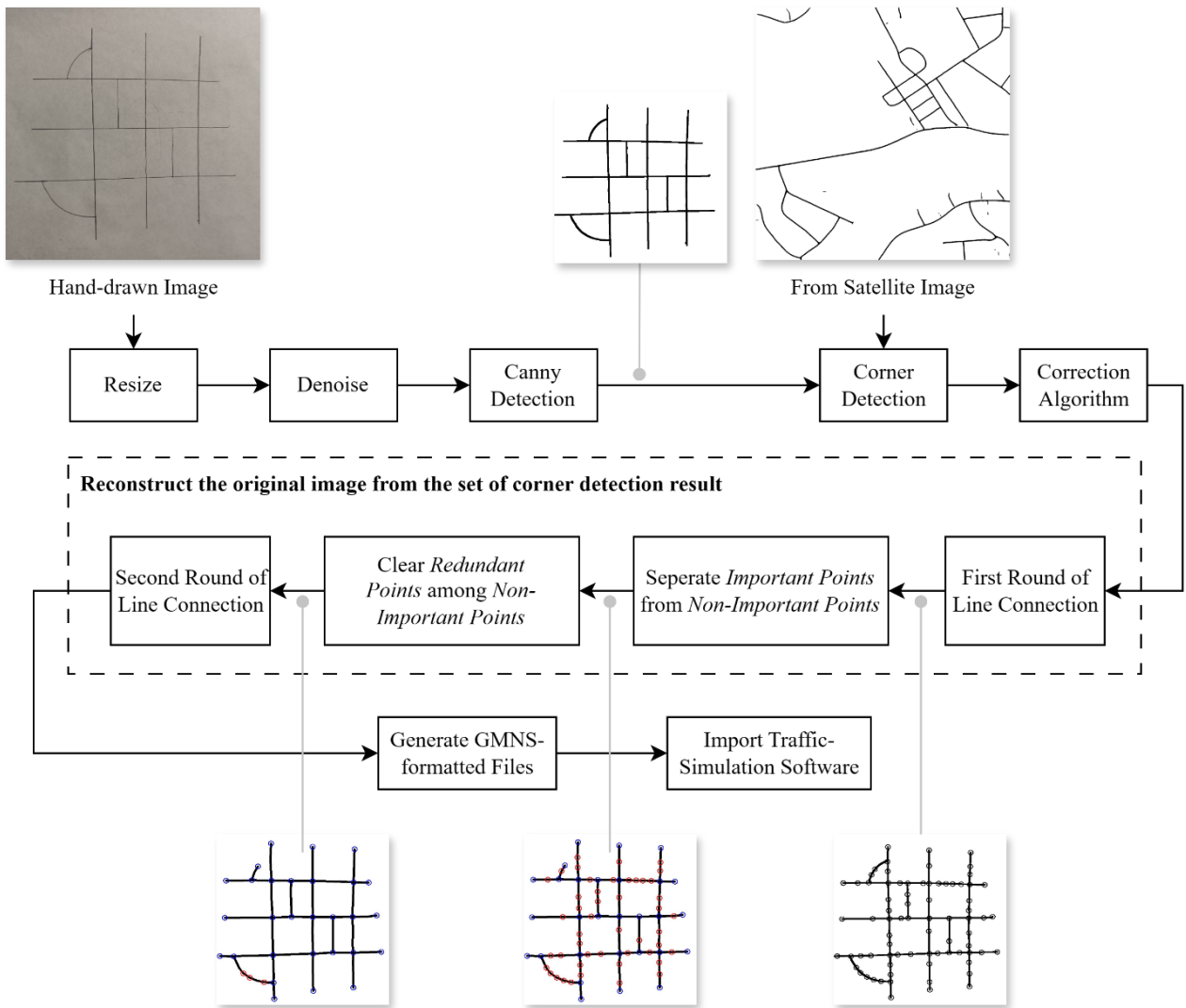

Figure 7. The overall structure of Image-based Road Network Generation

Shi-Tomasi corner detection is an improved version of the Harris corner detection algorithm. Shi-Tomasi found that the stability of corner detection is related to the smaller eigenvalue of the matrix M:

$$R = \min(\lambda_1, \lambda_2)$$

where R is the response function of the Harris corner detection algorithm. Using the smaller eigenvalue directly as the score avoids the need to adjust the sensitivity parameter K for corner detection. The specific details of the algorithm are not extensively elaborated in this paper.

The corner detection process results in numerous redundant points. However, during the actual simulation, only information about intersections, endpoints, and points on curves is required, and excess points can be inconvenient. The removal of redundant points can be achieved through the "Reconstruct the original image from the set of corner detection results" section in Figure 7.

- First Round of Line Connection Between Two Points

    The tool employs mean-square error for comparing the connected image with the original binary image. If the connection is correct, the error value decreases. Otherwise, if the line should not exist, the error value increases. This process iterates through all points obtained from corner detection, gathering coordinates of all connected points and information about each point's connection count.

- Filtering Based on Connection Count

    *Def: Points on intersections and endpoints are "Important Points", while the rest are "Non-Important Points".*

Assuming correct detection results, points connected only once are considered endpoints ("Important Points"), and points connected three times or more cannot be "Non-Important Points." Therefore, the tool categorizes points with connection counts not equal to 2 as "Important Points" and those with a count of 2 as "Non-Important Points."

- Filtering Based on Relative Position of Points

  *Def: If a "Non-Important Point" is located between two "Important Points," it is considered a redundant point.* Using this principle, the tool identifies a point as redundant if the distances between a "Non-important Point" and two "Important Points" are below a certain threshold. This step helps remove plenty of redundant points, especially those caused by recognition errors between two intersections.

- Second Round of Line Connection Between Two Points

  Following the acquisition of the final set of points, the tool undergoes an additional traversal for connecting points to generate new sets of nodes and lines, including their connection details. Subsequently, the data is stored in GMNS format. In the Node.csv file, the tool preserves the identifier and x, y coordinates of each point in the image. In the Link.csv file, the tool retains the identifier, corresponding identifiers of the two connected points, the length of the line, and the geographical coordinates transformed from x-y coordinates.

Upon obtaining the GMNS-formatted files, users can utilize Python's built-in xml.etree.ElementTree library to extract and convert point and line information into corresponding NOD.XML and EDG.XML files, facilitating further transformation. Finally, using SUMO's built-in Netconvert plugin, the acquired NOD.XML and EDG.XML files are combined and transformed into a road network file (NET.XML). This file encompasses information on points, lines, connections, and more, enabling direct importation into SUMO-related software.

**3.4.2 Road Network Generation from Hand-drown Image**

The process of generating a road network from a hand-drawn image is essentially similar to that from a satellite image, except that there is no need for U-Net to extract the road portion. As illustrated in Figure 7, the conversion of a hand-drawn image into a binary image initially involves dimension adjustment using the PIL library, followed by denoising, edge detection, and filling using the OpenCV library. Subsequent operations align with the previously described procedures.

**4 Case study**

In this section, we employ the ChatGPT model interface from OpenAI to implement multimodal road network generation across multiple computers with different configurations. To demonstrate the integrated functionalities of NGAI, including text-based road network generation (regular/location-specified road network generation) and image-based road network generation (road network generation from satellite/hand-drawn image), we provide one to two case studies for each functionality, showcasing them from the following four aspects.

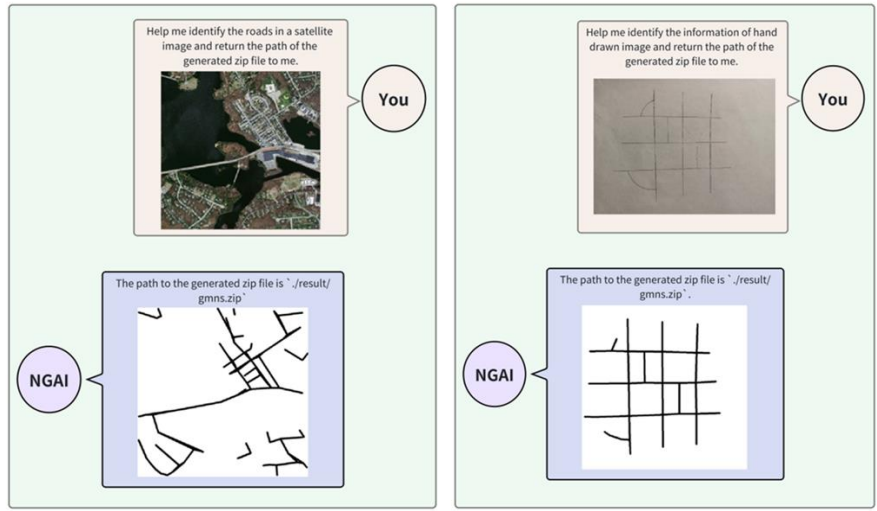

Figure 8. Application examples of NGAI

**4.1 Basic Capabilities**

As an AI traffic assistant, NGAI's basic capability lies in engaging in multi-turn dialogue based on instructions provided by human users and executing basic tasks. As depicted in Figure 8, it illustrates NGAI's fundamental ability to perform road network generation-related tasks based on natural language understanding. It is noteworthy that NGAI does not directly feed users' requests or provided images to LLM but instead offers a series of plugins available for road network generation for LLM to choose from. By judiciously selecting plugins, NGAI accurately understands users' demands, extracts information provided by users, and generates corresponding road network files to return to users.

**4.2 Accurate Road Network Identification and Generation**

To effectively provide users with accurate road network files, reducing the cumbersome work of users in road network drawing, NGAI must possess sufficiently accurate road network identification and drawing capabilities. For regular road network generation and location-specified road network generation, the accuracy of road network identification can be well ensured. For computer vision-based road network modeling, depending on the differences in satellite images or hand-drawn images to be processed, the results may have varying degrees of error. However, under the premise of ensuring low complexity of the road network, its accuracy can also be guaranteed to a certain extent, as illustrated in Figure 8. It is worth emphasizing that NGAI cannot autonomously modify users' requests and corresponding data but strictly provides services according to users' requirements, ensuring the reliability of road network files.

**4.3 Visualization of Road Networks**

As a human rather than a machine, users find it challenging to discern all information about road networks directly from road network files under the General Modeling Network Specification (GMNS) format. To address this issue, alongside providing road network files to users, NGAI also offers a corresponding road network overview map. As illustrated in Figure 8, in the overview map, users can readily comprehend the overall configuration of the road network as well as the relative relationships among its various components., thereby making basic judgments on whether the road network meets their needs.

**4.4 Fewer Error Occurrences**

Taking ChatGPT as an example, large language models possess the ability of self-reflection. During the operation of NGAI, after each plugin invocation, the LLM undergoes a thought process. Limited by the learning capabilities of the LLM itself, errors may occur at this stage, such as inaccurate plugin selection, repeated plugin invocation, or even directly considering the task cannot be completed. Therefore, we have provided detailed specifications and guidance for LLM within the plugin descriptor, prohibiting it from fabricating any data. Through multiple experiments, the descriptions of each plugin are optimized to minimize the occurrence of errors by the LLM.

We conducted experiments over 10 times for each plugin, considering different languages (Chinese, English) and different input forms of questions (detailed questions, concise questions, questions with only keywords), yielding the following data.

Table 1. Experiment Result

| Indicator | Tool selection accuracy | | Average invocation count | | Repeated call probability | |
|---|---|---|---|---|---|---|
| Expression form \ Language | English | Chinese | English | Chinese | English | Chinese |
| Detailed | 1.0000 | 1.0000 | 1.0500 | 1.0000 | 0.0050 | 0.0000 |
| Concise | 0.9250 | 0.9750 | 1.0818 | 1.0000 | 0.0541 | 0.0000 |
| Keywords | 0.8000 | 1.0000 | 1.6563 | 1.3750 | 0.3750 | 0.3500 |
| Indicator comprehensive value | 0.9083 | 0.9917 | 1.2385 | 1.1261 | 0.1468 | 0.1176 |
| Overall indicator comprehensive value | 0.9500 | | 1.1795 | | 0.1316 | |

From the table, we can clearly understand the interaction between LLM and road network modeling plugins from three dimensions: tool selection accuracy, average invocation count, and repeated call probability. Taking Chinese questions as an example, we can ascertain that LLM has a very high probability of extracting the correct tool, which is 0.9917. Meanwhile, the occurrence of LLM's repeated invocation of tools is extremely low, apart from keywords input, the average invocation count is 1.0000, and the Repeated call probability is 0.0000. LLM hardly repeats calling tools, ensuring the rapidity and accuracy of NGAI's operation. However, for questions that only provide keywords, LLM may sometimes struggle to understand users' demands, leading to the occurrence of repeated invocation due to misjudgment. Nonetheless, providing as detailed questions as possible can mitigate this issue. Unexpectedly, although NGAI is coded entirely in English, performance metrics exhibit enhancement when employing Chinese prompts, outperforming those attained with English prompts. With Chinese prompts, NGAI achieves a tool selection accuracy of 99.17%, whereas under English prompts, it registers 90.83%. The repeated call probability under Chinese prompts, for both detailed and concise descriptions, is 0, while the tool selection accuracy under keyword prompts reaches 100.00%. This occurrence could be ascribed to the inherent attributes of LLMs or disparities in the linguistic attributes between Chinese and English.

In summary, this section demonstrates NGAI's efficient handling of road network modeling tasks under various demands from users, significantly reducing the workload of users in road network modeling and showcasing the advantages of multimodal road network generation based on LLMs in terms of easy operation and high efficiency.

**5 Conclusion**

This paper introduces a model that simplifies the road network modeling steps in transportation simulation. Leveraging the logical reasoning and multimodal recognition capabilities of LLMs, NGAI integrates four road network modeling plugins to achieve automated road network modeling that supports multimodal data input. The results of the case study indicate that when posing detailed questions in English to NGAI, the tool selection accuracy reaches 100.00%, with a repeat invocation probability of 0.50%. Furthermore, the results suggest that NGAI performs even better when using Chinese, with superior tool selection accuracy, average invocation count, and repeat invocation probability compared to English.

The main contributions of this paper are as follows. The research proposes a multimodal, low-cost, and accessible form of road network modeling, capable of generating road networks based on demand while ensuring sufficient precision. This integration revolutionizes the approach to traffic simulation, optimizing the steps users take when utilizing simulation software, thus rendering traffic simulation tasks simpler and more intelligent. Additionally, this study demonstrates the formidable logical reasoning capabilities and recognition abilities of multimodal information inherent in LLMs, as well as their capacity to resolve user issues with appropriate prompts. The integration of LLM functionality with road network modeling plugins offers a novel perspective for the application of LLMs in the transportation domain, pointing to a new direction for the application of artificial intelligence in this field.

However, there still exist some limitations to the proposed model. Firstly, the accuracy of satellite image recognition depends on the quality of the satellite images in the training set and the source of the satellite images provided by users. Especially for images with poor quality, the recognition accuracy may be compromised. Therefore, future efforts will focus on increasing the training volume of the model to improve the accuracy of satellite image recognition. Additionally, while NGAI performs well in response to Chinese questions, its response to English questions is relatively poorer, with instances of repeated invocations. Future research will investigate the reasons for this phenomenon and explore potential improvements through training specialized language models or utilizing other large language models.